\documentclass[12pt]{article}
\usepackage{amsmath}
\usepackage{amssymb}
\usepackage{amsfonts}

\newcommand{\field}[1]{\mathbb{#1}}
\newcommand{\R}{\field{R}}
\newcommand{\crr}{{\cal R}}
\newcommand{\cii}{{\cal I}}
\newcommand{\chh}{{\cal H}}
\newcommand{\cxx}{{\cal X}}
\newcommand{\cyy}{{\cal Y}}
\newcommand{\bp}{b^{\dagger}}
\newcommand{\bb}{\bar{b}}
\newcommand{\bbp}{\bar{b}^{\dagger}}

\title{Some Properties of an Infinite Family of Deformations of the Harmonic Oscillator}
\author{Christiane Quesne\\ 
{\small Physique Nucl\'eaire Th\'eorique et Physique Math\'ematique,  Universit\'e Libre de Bruxelles,} \\ 
{\small Campus de la Plaine CP229, Boulevard~du Triomphe, B-1050 Brussels, Belgium}}
\date{ }
\begin{document}
\baselineskip=22pt plus 1pt minus 1pt
%%%%%%%%%%%%%%%%%%%%%%%%%%%%%%%%%%%%%%%%%%%%%%%%%%%%%%%%%%
\maketitle

\begin{abstract}
In memory of Marcos Moshinsky, who promoted the algebraic study of the harmonic oscillator, some results recently obtained on an infinite family of deformations of such a system are reviewed. This set, which was introduced by Tremblay, Turbiner, and Winternitz, consists in some Hamiltonians $H_k$ on the plane, depending on a positive real parameter $k$. Two algebraic extensions of $H_k$ are described. The first one, based on the elements of the dihedral group $D_{2k}$ and a Dunkl operator formalism, provides a convenient tool to prove the superintegrability of $H_k$ for odd integer $k$. The second one, employing two pairs of fermionic operators, leads to a supersymmetric extension of $H_k$ of the same kind as the familiar Freedman and Mende super-Calogero model. Some connection between both extensions is also outlined.
\end{abstract}


%
%========================================================================
%
\newpage
\section{INTRODUCTION}

During his long and prolific career, the study of the harmonic oscillator has been one of the prominent topics dealt with by Marcos Moshinsky \cite{brody, moshinsky96}, with whom I have enjoyed the privilege of collaborating for many years including on this subject \cite{moshinsky74, cq90, moshinsky95}. As a token of reminiscence and gratitude for all what I learnt from him, in this paper dedicated to his memory I will review some recent results on a related problem.\par
%
%---------------------------------------------------------------------------------------------------
%
In \cite{tremblay}, Tremblay, Turbiner and Winternitz (TTW) indeed introduced an infinite family of exactly solvable Hamiltonians
\begin{equation}
\begin{split}
  & H_k = - \partial_r^2 - \frac{1}{r} \partial_r - \frac{1}{r^2} \partial_{\varphi}^2 + \omega^2 r^2 +
       \frac{k^2}{r^2} [a(a-1) \sec^2 k\varphi + b(b-1) \csc^2 k\varphi], \\
  & 0 \le r < \infty, \qquad 0 \le \varphi < \frac{\pi}{2k},
\end{split}
\end{equation}
which may be considered as deformations of the harmonic oscillator on a plane and reduce for $k=1$, 2, 3 to those of the familiar Smorodinsky-Winternitz \cite{fris, winternitz}, $BC_2$ model \cite{olshanetsky} and Calogero-Marchioro-Wolfes \cite{wolfes, calogero} systems. They showed that for any real $k$, $H_k$ is integrable with $X_k = - \partial_{\varphi}^2 + k^2 [a(a-1) \sec^2 k\varphi + b(b-1) \csc^2 k\varphi]$ an integral of motion. They also conjectured (and actually proved for $k=1$, 2, 3, 4) that it is superintegrable for any integer $k$, the second integral of motion $Y_{2k}$ being some $2k$th-order differential operator. Later on, this was established  by myself for odd $k$ \cite{cq10a}, then by Kalnins, Kress and Miller for integer (or even rational) $k$ \cite{kalnins}.   \par
%
%---------------------------------------------------------------------------------------------------------
%
In the following, two extensions of $H_k$ will be considered. The first one \cite{cq10b}, based on the elements of the dihedral group $D_{2k}$, was used in the superintegrability proof of Ref.\ \cite{cq10a} while the second one \cite{cq10c}, employing two pairs of fermionic operators, led to a supersymmetric extension of the same kind as the familiar Freedman and Mende super-Calogero model \cite{freedman}. Some connection between both extensions \cite{cq10d} will also be reviewed.\par
%
%=========================================================================
%
\section{DIHEDRAL GROUP EXTENSION}

The dihedral group $D_{2k}$ has $4k$ elements $\crr^i$ and $\crr^i \cii$, $i=0$, 1, \ldots, $2k-1$, satisfying the relations 
\begin{equation}
  \crr^{2k} = \cii^2 = 1, \qquad \cii \crr = \crr^{2k-1} \cii, \qquad \crr^{\dagger} = \crr^{2k-1}, \qquad 
  \cii^{\dagger} = \cii. 
\end{equation}
They are realizable on the Euclidean plane as the rotation operator through angle $\pi/k$, $\crr = \exp\left(\frac{1}{k} \pi \partial_{\varphi}\right)$, and the operator changing $\varphi$ into $-\varphi$, $\cii = \exp({\rm i} \pi \varphi \partial_{\varphi})$ \cite{cq10b}.\par
%
%-------------------------------------------------------------------------------------------------------------------
%
With their use, one can extend the partial derivatives $\partial_r$ and $\partial_{\varphi}$ into some differential-difference operators,
\begin{equation}
\begin{split}
  D_r &= \partial_r - \frac{1}{r} (a\crr + b) \left(\sum_{i=0}^{k-1} \crr^{2i}\right) \cii, \\
  D_{\varphi} &= \partial_{\varphi} + a \sum_{i=0}^{k-1} \tan\left(\varphi + \frac{i\pi}{k}\right) \crr^{k+2i} \cii
        - b \sum_{i=0}^{k-1} \cot\left(\varphi + \frac{i\pi}{k}\right) \crr^{2i} \cii,
\end{split} \label{eq:D}  
\end{equation}
similar to Dunkl operators \cite{dunkl}. They satisfy more complicated relations than $\partial_r$ and $\partial_{\varphi}$, namely
\begin{equation}
\begin{split}
  & D_r^{\dagger} = - D_r - \frac{1}{r} \left[1 + 2(a\crr + b) \left(\sum_{i=0}^{k-1} \crr^{2i} \right) \cii\right], 
          \qquad D_{\varphi}^{\dagger} = - D_{\varphi}, \\  
  & \crr D_r = D_r \crr, \qquad \cii D_r = D_r \cii, \qquad \crr D_{\varphi} = D_{\varphi} \crr, \qquad 
          \cii D_{\varphi} = - D_{\varphi} \cii, \\
  & [D_r, D_{\varphi}] = - \frac{2}{r} (a\crr + b) \left(\sum_{i=0}^{k-1} \crr^{2i} \right) \cii D_{\varphi}.
\end{split} 
\end{equation}
In (\ref{eq:D}), $k$ may be any odd integer. For even $k$, $D_r$ and $D_{\varphi}$ assume a different form, which we will not consider here.\par
%
%----------------------------------------------------------------------------------------------------------------------------------
%   
The TTW Hamiltonian $H_k$ and its integral of motion $X_k$ can be generalized by incorporating the elements of $D_{2k}$, as well as the new operators $D_r$ and $D_{\varphi}$. The resulting $D_{2k}$-extended operators
\begin{equation}
\begin{split}
  \chh_k & = - D_r^2 - \frac{1}{r} \left[1 + 2 (a\crr + b) \left(\sum_{i=0}^{k-1} \crr^{2i}\right) \cii\right] D_r
  - \frac{1}{r^2} D_{\varphi}^2 + \omega^2 r^2  \\ 
  & = - \partial_r^2 - \frac{1}{r} \partial_r - \frac{1}{r^2} \left[D_{\varphi}^2 - k (a^2 + b^2 + 2ab \crr)
  \sum_{i=0}^{k-1} \crr^{2i}\right] + \omega^2 r^2
\end{split} \label{eq:dihedral-H}
\end{equation}
and
\begin{equation}
\begin{split}
  \cxx_k & = - D_{\varphi}^2 = - \partial_{\varphi}^2 + \sum_{i=0}^{k-1} \sec^2 \left(\varphi 
       + \frac{i\pi}{k}\right) a (a - \crr^{k+2i} \cii)  \\
  & \quad + \sum_{i=0}^{k-1} \csc^2 \left(\varphi + \frac{i\pi}{k}\right) b (b - \crr^{2i} \cii) 
       - k (a^2 + b^2 + 2ab \crr) \sum_{i=0}^{k-1} \crr^{2i}  
\end{split}
\end{equation}
turn out to be invariant under $D_{2k}$ and to give back $H_k$ and $X_k$ after projection in the $D_{2k}$ identity representation, i.e., by letting $\crr$ and $\cii$ go to 1.\par
%
%++++++++++++++++++++++++++++++++++++++++++++++++++++++++++++++++++++
%
\subsection{\boldmath Application to the superintegrability problem of $H_k$ for odd $k$}

The formalism considered above can still be enlarged \cite{cq10a} by introducing a set of $k$ (dependent) pairs of modified boson creation and annihilation operators
\begin{equation}
\begin{split}
  A_i &= \frac{1}{\sqrt{2\omega}} \left[\cos \left(\varphi + \frac{i\pi}{k}\right) (\omega r + D_r) - \frac{1}{r}
       \sin\left(\varphi + \frac{i\pi}{k}\right) D_{\varphi}\right], \\
  A_i^{\dagger} &= \frac{1}{\sqrt{2\omega}} \left[\cos \left(\varphi + \frac{i\pi}{k}\right) (\omega r - D_r) 
       + \frac{1}{r} \sin\left(\varphi + \frac{i\pi}{k}\right) D_{\varphi}\right],       
\end{split}  
\end{equation}
where $i=0$, 1, \ldots,~$k-1$. One can show that $A_i^{\dagger}$ is the Hermitian conjugate of $A_i$ and that the $2k$ operators $A_i$ and $A_i^{\dagger}$ satisfy the modified commutations relations
\begin{equation}
\begin{split}
  [A_i, A_j] & = [A_i^{\dagger}, A_j^{\dagger}] = 0, \\ 
  [A_i, A_j^{\dagger}] &= [A_j, A_i^{\dagger}] = \cos \frac{(j-i)\pi}{k} + 2a \sum_l \cos \frac{(l-i)\pi}{k}
      \cos \frac{(l-j)\pi}{k} \crr^{k+2l} \cii  \\
  & \quad + 2b \sum_l \sin \frac{(l-i)\pi}{k} \sin \frac{(l-j)\pi}{k} \crr^{2l} \cii,
\end{split}  
\end{equation}
for $i, j=0$, 1, \ldots,~$k-1$. Here all summations over $l$ run from 0 to $k-1$. The $A_i$ also fulfil the exchange relations
\begin{equation}
\begin{split}
  \crr A_i \crr^{-1} &= A_{i+1}, \qquad i = 0, 1, \ldots, k-2, \qquad \crr A_{k-1} \crr^{-1} = - A_0, \\
  \cii A_0 \cii^{-1} & = A_0, \qquad \cii A_i \cii^{-1} = - A_{k-i}, \qquad i = 1, 2, \ldots, k-1. 
\end{split}  
\end{equation}
\par
%
%----------------------------------------------------------------------------------------------------------------------
%
These modified boson operators can be used to define $k$ modified oscillator Hamiltonians
\begin{equation}
  H_i = \tfrac{1}{2} \{A_i^{\dagger}, A_i\}, \qquad i = 0, 1, \ldots, k-1,
\end{equation}
which transform among themselves under $D_{2k}$:
\begin{equation}
\begin{split}
 \crr H_i \crr^{-1} &= H_{i+1}, \qquad i = 0, 1, \ldots, k-2, \qquad \crr H_{k-1} \crr^{-1} = H_0,  \\
 \cii H_0 \cii^{-1} &= H_0, \qquad \cii H_i \cii^{-1} = H_{k-i}, \qquad i = 1, 2, \ldots, k-1. 
\end{split}  \label{eq:D-H}
\end{equation}
\par
%
%----------------------------------------------------------------------------------------------------------------------
%
{}From the explicit expressions of the $H_i$'s, namely
\begin{equation}
\begin{split}
  & 2\omega H_i = - \cos^2 \biggl(\varphi + \frac{i\pi}{k}\biggr) D_r^2 + \frac{1}{r} \sin \biggl(\varphi + 
      \frac{i\pi}{k}\biggr) \cos \biggl(\varphi + \frac{i\pi}{k}\biggr) (D_r D_{\varphi} + D_{\varphi} D_r) \\
  & \quad - \frac{1}{r^2} \sin^2 \biggl(\varphi + \frac{i\pi}{k}\biggr) D_{\varphi}^2 \\
  & \quad - \frac{1}{r} \biggl[\sin^2 \biggl(\varphi + \frac{i\pi}{k}\biggr) + 2a \sum_l \cos^2 \frac{(l-i)\pi}{k}
      \crr^{k+2l} \cii + 2b \sum_l \sin^2 \frac{(l-i)\pi}{k} \crr^{2l} \cii\biggr] D_r \\
  & \quad + \frac{1}{r^2} \biggl[- 2 \sin \biggl(\varphi + \frac{i\pi}{k}\biggr) \cos \biggl(\varphi + \frac{i\pi}{k}
      \biggr) - 2a \sum_l \sin \frac{(l-i)\pi}{k} \cos \frac{(l-i)\pi}{k} \crr^{k+2l} \cii \\
  & \quad + 2b \sum_l \sin \frac{(l-i)\pi}{k} \cos \frac{(l-i)\pi}{k} \crr^{2l} \cii\biggr] D_{\varphi} + \omega^2
      r^2 \cos^2 \biggl(\varphi + \frac{i\pi}{k}\biggr),   
\end{split}
\end{equation}
it can be shown that such operators are connected with the $D_{2k}$-extended Hamiltonian through the relation
\begin{equation}
  2\omega \sum_{i=0}^{k-1} H_i = \frac{k}{2} \chh_k.  
\end{equation}
Hence $\chh_k$ may be considered as a modified boson oscillator Hamiltonian.\par
%
%-------------------------------------------------------------------------------------------------------------------
%
Next, it can be proved that all the $H_i$'s commute with $\chh_k$ and are therefore integrals of motion for the latter. From them, one can form two $D_{2k}$ invariants, namely their sum proportional to $\chh_k$ and their symmetrized product
\begin{equation}
  \cyy_{2k} = (2\omega)^k \sum_p H_{p(0)} H_{p(1)} \cdots H_{p(k-1)},  
\end{equation}
where the summation runs over all $k!$ permutations of 0, 1, \ldots,~$k-1$.\par
%
%---------------------------------------------------------------------------------------------------------------
%
Projection in the $D_{2k}$ identity representation then leads to $H_k$, on one hand, and to an integral of motion $Y_{2k}$ of the latter, on the other hand. From its definition, it is clear that $Y_{2k}$ is a differential operator of order $2k$. It is also functionally independent of $X_k$ because it can be established that one of the highest-order terms in $[\cxx_k, \cyy_{2k}]$ does not vanish and remains nonvanishing after projection in the $D_{2k}$ identity representation, thereby proving that $[X_k, Y_{2k}] \ne 0$.\par
%
%--------------------------------------------------------------------------------------------------------
%
We conclude that for any odd $k$, the operators $X_k$ and $Y_{2k}$ provide us with a set of two functionally independent integrals of motion of $H_k$, which is therefore superintegrable as claimed in the TTW conjecture.\par
%
%==========================================================
%
\section{\boldmath ${\cal N}=2$ supersymmetric extension}

By using two independent pairs of fermionic creation and annihilation operators $(\bp_x, b_x)$ and $(\bp_y, b_y)$, the TTW Hamiltonian $H_k$ can be extended into a supersymmetric Hamiltonian $\chh^s$ \cite{cq10c}. This can be carried out in the framework of an $osp(2/2, \R)$ superalgebra with even generators $K_0$, $K_{\pm}$, $Y$ (closing the $sp(2, \R) \times so(2)$ Lie algebra) and odd generators $V_{\pm}$, $W_{\pm}$ (which are two $sp(2, \R)$ spinors). The corresponding (nonvanishing) commutation or anticommutation relations and Hermiticity properties are given by
\begin{equation}
\begin{array}{ll}
  [K_0, K_{\pm}] = \pm K_{\pm}, &\qquad  [K_+, K_-] = - 2K_0, \\[0.3cm]
  [K_0, V_{\pm}] = \pm \tfrac{1}{2} V_{\pm}, &\qquad [K_0, W_{\pm}] = \pm \tfrac{1}{2} W_{\pm}, 
  \\[0.3cm]
  [K_{\pm}, V_{\mp}] = \mp V_{\pm}, &\qquad [K_{\pm}, W_{\mp}] = \mp W_{\pm}, \\[0.3cm]
  [Y, V_{\pm}] = \tfrac{1}{2} V_{\pm}, &\qquad [Y, W_{\pm}] = - \tfrac{1}{2} W_{\pm}, \\[0.3cm] 
  \{V_{\pm}, W_{\pm}\} = K_{\pm}, &\qquad \{V_{\pm}, W_{\mp}\} = K_0 \mp Y      
\end{array}  
\end{equation}
and
\begin{equation}
  K_0^{\dagger} = K_0, \qquad K_{\pm}^{\dagger} = K_{\mp}, \qquad Y^{\dagger} = Y, \qquad 
  V_{\pm}^{\dagger} = W_{\mp},
\end{equation}
respectively.\par
%
%------------------------------------------------------------------------------------------------------------
%
Standard supersymmetric quantum mechanics \cite{cooper}, with supersymmetric Hamiltonian ${\cal H}^s$ and supercharges $Q$, $Q^{\dagger}$ such that
\begin{equation}
  [{\cal H}^s, Q] = [{\cal H}^s, Q^{\dagger}] = 0, \qquad \{Q, Q^{\dagger}\} = {\cal H}^s, 
\end{equation}
is realized by the three operators
\begin{equation}
  {\cal H}^s = 4 \omega (K_0 + Y), \qquad Q = 2 \sqrt{\omega}\, W_+, \qquad Q^{\dagger} = 
  2 \sqrt{\omega}\, V_-  
\end{equation}
generating a $sl(1/1)$ subsuperalgebra of $osp(2/2, \R)$.\par
%
%----------------------------------------------------------------------------------------------------------
%  
To get simpler expressions of the generators, it is useful to introduce new `rotated' fermionic operators defined by
\begin{equation}
  \bbp_x = \bp_x \cos \varphi + \bp_y \sin \varphi, \qquad \bbp_y = - \bp_x \sin \varphi + \bp_y \cos \varphi
\end{equation}
and similarly for $\bb_x$, $\bb_y$. Such a transformation, however, breaks the commutativity of bosonic and fermionic degrees of freedom (e.g., $[\partial_{\varphi}, \bb_x] = \bb_y$).\par
%
%-------------------------------------------------------------------------------------------------------
%
The resulting even and odd $osp(2/2, \R)$ generators can be expressed as
\begin{equation}
\begin{split}
  & K_0  = K_{0,{\rm B}} + \Gamma, \qquad K_{\pm} = K_{\pm,{\rm B}} - \Gamma, \\
  & K_{0,{\rm B}} = \frac{1}{4 \omega} H_k, \qquad K_{\pm,{\rm B}} = \frac{1}{4 \omega} [- H_k 
         + 2 \omega^2 r^2 \mp 2 \omega (r \partial_r + 1)] \\
  & \Gamma = \frac{k}{2\omega r^2} \bigl\{a \bigl[\bbp_x \bb_x - \tan k\varphi \bigl(\bbp_x \bb_y 
       + \bbp_y \bb_x\bigr) + (k \sec^2 k\varphi - 1) \bbp_y \bb_y\bigr] \\
  & \quad + b \bigl[\bbp_x \bb_x + \cot k\varphi \bigl(\bbp_x \bb_y + \bbp_y \bb_x\bigr)
       + (k \csc^2 k\varphi - 1) \bbp_y \bb_y\bigr]\bigr\}, \\ 
  & Y = \frac{1}{2} \bigl[\bbp_x \bb_x + \bbp_y \bb_y - k(a+b) - 1\bigr]
\end{split}
\end{equation}
and
\begin{equation}
\begin{split}
  V_{\pm} & = \frac{1}{2\sqrt{\omega}} \biggl[\bbp_x \biggl(\mp \partial_r + \omega r \pm \frac{k(a+b)}{r}
         \biggr) \mp \bbp_y \frac{1}{r} (\partial_{\varphi} + ka \tan k\varphi - kb \cot k\varphi)\biggr], \\
  W_{\pm} & = \frac{1}{2\sqrt{\omega}} \biggl[\bb_x \biggl(\mp \partial_r + \omega r \mp \frac{k(a+b)}{r}
         \biggr) \mp \bb_y \frac{1}{r} (\partial_{\varphi} - ka \tan k\varphi + kb \cot k\varphi)\biggr],
\end{split}  
\end{equation}
respectively.\par
%
%------------------------------------------------------------------------------------------------------------------
%
On starting from the wavefunctions of the TTW Hamiltonian $H_k$
\begin{equation}
\begin{split}
  & \Psi_{N,n}(r, \varphi) = {\cal N}_{N,n} Z^{(2n+a+b)}_N (z) \Phi^{(a,b)}_n (\varphi), \\
  & Z^{(2n+a+b)}_N (z) = \left(\frac{z}{\omega}\right)^{\left(n + \frac{a+b}{2}\right) k} L^{((2n+a+b)k)}_N
          (z) e^{- \frac{1}{2} z}, \qquad z = \omega r^2, \\
  & \Phi^{(a,b)}_n (\varphi) = \cos^a k\varphi \sin^b k\varphi P^{\left(a - \frac{1}{2}, b - \frac{1}{2}\right)}_n
          (\xi), \qquad \xi = - \cos 2k\varphi, \\[0.2cm]
  & N, n = 0, 1, 2, \ldots,
\end{split}
\end{equation}
defined in terms of Laguerre and Jacobi polynomials and such that 
\begin{equation}
\begin{split}
  & H_k \Psi_{N,n}(r, \varphi) = E_{N,n} \Psi_{N,n}(r, \varphi), \qquad E_{N,n} = 2\omega 
        [2N + (2n+a+b)k + 1], \\
  & \int_0^{\infty} dr\, r \int_0^{\pi/(2k)} d\varphi\, |\Psi_{N,n}(r, \varphi)|^2 = 1,
\end{split}
\end{equation}
one gets eigenstates of the supersymmetrized TTW Hamiltonian $\chh^{s}$ after multiplication by the fermionic vacuum state $|0\rangle$. The corresponding eigenvalues are ${\cal E}_{N,n} = E_{N,n} - E_{0,0} = 4\omega (N + nk)$. Such extended states are also eigenstates of the $osp(2/2, \R)$ weight generators $K_0$ and $Y$, corresponding to the eigenvalues
\begin{equation}
  \tau = \left(n + \frac{a+b}{2}\right)k + \frac{1}{2}, \qquad q = - \frac{1}{2}[(a+b)k + 1],
\end{equation}
respectively.\par
%
%------------------------------------------------------------------------------------------------------------------
% 
{}For each value of $n \in \{0, 1, 2, \ldots\}$ (specifying the angular wavefunctions of $H_k$ as well as the eigenvalues of the first integral of motion $X_k$), it is possible to construct an $osp(2/2, \R)$ irreducible representation (irrep) characterized by $(\tau, q)$. Its nature, however, depends on the value assumed by $n$.\par
%
%------------------------------------------------------------------------------------------------------------------
%
{}For $n=0$, one obtains a lowest-weight state (LWS) irrep based on the extended ground state $\Psi_{0,0}(r, \varphi) |0\rangle$. This state is indeed annihilated by all the lowering generators $K_-$, $V_-$, $W_-$ of $osp(2/2, \R)$. The irrep is a so-called atypical one with $\tau = - q$ (which means that the vanishing Casimir operators $C_2$ and $C_3$ cannot specify the irrep). It contains only two $sp(2, \R) \times so(2)$ irreps: $(\tau) (q)$ and $(\tau + \frac{1}{2}) (q + \frac{1}{2})$ (spanned by zero- and one-fermion states, respectively).\par
%
%--------------------------------------------------------------------------------------------------------------
% 
{}For any $n \ne 0$, one gets an $osp(2/2, \R)$ irrep containing four $sp(2, \R) \times so(2)$ irreps: $(\tau) (q)$, $(\tau - \frac{1}{2}) (q + \frac{1}{2})$, $(\tau + \frac{1}{2}) (q + \frac{1}{2})$, and $(\tau) (q+1)$. The first one is spanned by zero-fermion states, the next two by a mixture of one-fermion states and the last one by two-fermion states. No state is annihilated by all the $osp(2/2, \R)$ lowering generators.\par
%
%------------------------------------------------------------------------------------------------------------
%
The eigenvalues of the two Casimir operators of $osp(2/2, \R)$ are given by
\begin{equation}
  C_2 \to n (n+a+b) k^2, \qquad C_3 \to - \tfrac{1}{2} (a+b) n (n+a+b) k^2,
\end{equation}    
which proves the above-mentioned result for $n=0$.\par
%
%-------------------------------------------------------------------------------------------------------
%
The supersymmetric extension presented in this section is valid for any real value of $k$. For the special cases of $k=1$, 2, 3, it gives back some known results related to the super-Calogero model \cite{freedman} and to the supersymmetrization of other Calogero-like systems \cite{brink}.\par
%
%==========================================================
%
\section{CONNECTION BETWEEN BOTH EXTENSIONS}

To start with, it is possible to realize the elements $\crr^i$ and $\crr^i \cii$, $i=0$, 1, \ldots, $2k-1$, of the dihedral group $D_{2k}$ in terms of two independent pairs of fermionic operators $(\bp_x, b_x)$ and $(\bp_y, b_y)$ \cite{cq10d}. On starting from the definitions
\begin{equation}
\begin{split}
  \crr & \equiv 1 + \left(\cos \frac{\pi}{k} - 1\right) (\bp_x b_x + \bp_y b_y) + \sin \frac{\pi}{k} (\bp_x b_y
       - \bp_y b_x) \\
 & \quad + 2 \left(1 - \cos \frac{\pi}{k}\right) \bp_x b_x \bp_y b_y, \\
  \cii & \equiv 1 - 2 \bp_y b_y = - [\bp_y, b_y], 
\end{split} 
\end{equation}
one can indeed show that for any $i=0$, 1, \ldots,~$2k-1$ 
\begin{equation}
\begin{split}
  \crr^i & = 1 + \left(\cos \frac{i\pi}{k} - 1\right) (\bp_x b_x + \bp_y b_y) + \sin \frac{i\pi}{k} (\bp_x b_y
       - \bp_y b_x) \\
  & \quad + 2 \left(1 - \cos \frac{i\pi}{k}\right) \bp_x b_x \bp_y b_y,  \\
  \crr^i \cii & = 1 + \left(\cos \frac{i\pi}{k} - 1\right) \bp_x b_x - \left(\cos \frac{i\pi}{k} + 1\right) \bp_y b_y
       - \sin \frac{i\pi}{k} (\bp_x b_y + \bp_y b_x) 
\end{split}  \label{eq:substitution} 
\end{equation}
and that such operators satisfy all defining relations of $D_{2k}$.\par
%
%------------------------------------------------------------------------------------------------------------
%
The next step consists in making the substitution (\ref{eq:substitution}) in the $D_{2k}$-extended TTW Hamiltonian, given in (\ref{eq:dihedral-H}). As a result, the latter is mapped onto the difference between the supersymmetric TTW Hamiltonian $\chh^s$ and its purely fermionic term $4 \omega Y$ provided the trigonometric identities
\begin{equation}
\begin{split}
  & \sum_{i=0}^{k-1} \tan \left(\varphi + \frac{i\pi}{k}\right) \cos \frac{2i\pi}{k} = - k \frac{\sin[(k-2)\varphi]}
       {\cos k\varphi}, \\
  & \sum_{i=0}^{k-1} \tan \left(\varphi + \frac{i\pi}{k}\right) \sin \frac{2i\pi}{k} = k \frac{\cos[(k-2)\varphi]}
       {\cos k\varphi} - \delta_{k,1}
\end{split}  
\end{equation}
are satisfied. A simple proof of these relations has been found, thereby establishing a connection between the $D_{2k}$ and the supersymmetric extensions of the TTW Hamiltonian.\par
%
%=================================================================
%

\end{document}